\documentstyle[prl,aps,floats]{revtex}
\input epsf
\newcommand{\pbp}      {$\overline{p}p$} 
\newcommand{\invpb}    {pb$^{-1}$}
\newcommand{\apts}     {$\langle p_{T} \rangle^2$}
\newcommand{\qapts}    {$Q^2 = \langle p_{T} \rangle^2$}
\newcommand{\qhs}      {$M_{W}^{2} + p_{T_{W}}^{2}$}
\newcommand{\qqhs}     {$Q^2 = M_{W}^{2} + p_{T_{W}}^{2}$}
\newcommand{\et}       {$E_{T}$}
\newcommand{\mjj}      {$M_{jj}$}
\newcommand{\drjj}     {$\Delta R_{jj}$}
\newcommand{\Zee}      {$Z \rightarrow e^+e^-$}
\newcommand{\Wenu}     {$W^{\pm} \rightarrow e^{\pm}\nu$}
\newcommand{\met}      {\not\!\!\!{E}_{T}}
\font\eightit=cmti8
\def\r#1{\ignorespaces $^{#1}$}

\begin{document}
\draft
\title{Properties of Jets in $W$ Boson Events from 1.8 TeV \pbp\ Collisions}
\author{
\begin{sloppypar}
\noindent
F.~Abe,\r {17} H.~Akimoto,\r {38}
A.~Akopian,\r {31} M.~G.~Albrow,\r 7 S.~R.~Amendolia,\r {27} 
D.~Amidei,\r {20} J.~Antos,\r {33} S.~Aota,\r {36}
G.~Apollinari,\r {31} T.~Arisawa,\r {38}
% ...
\end{sloppypar}
\begin{center}
\vskip -0.1in
(CDF Collaboration)
\vskip -0.1in
\end{center}
}
\address{
\begin{center}
\r 1  {\eightit Argonne National Laboratory, Argonne, Illinois 60439} \\
\r 2  {\eightit Istituto Nazionale di Fisica Nucleare, University of Bologna,
I-40127 Bologna, Italy} \\
\r 3  {\eightit Brandeis University, Waltham, Massachusetts 02254} \\
\r 4  {\eightit University of California at Los Angeles, Los 
Angeles, California  90024} \\  
\r 5  {\eightit University of Chicago, Chicago, Illinois 60637} \\
\r 6  {\eightit Duke University, Durham, North Carolina  27708} \\
\r 7  {\eightit Fermi National Accelerator Laboratory, Batavia, Illinois 
60510} \\
\r 8  {\eightit University of Florida, Gainesville, FL  32611} \\
\r 9  {\eightit Laboratori Nazionali di Frascati, Istituto Nazionale di Fisica
               Nucleare, I-00044 Frascati, Italy} \\
\r {10} {\eightit University of Geneva, CH-1211 Geneva 4, Switzerland} \\
\r {11} {\eightit Harvard University, Cambridge, Massachusetts 02138} \\
\r {12} {\eightit Hiroshima University, Higashi-Hiroshima 724, Japan} \\
\r {13} {\eightit University of Illinois, Urbana, Illinois 61801} \\
\r {14} {\eightit Institute of Particle Physics, McGill University, Montreal 
H3A 2T8, and University of Toronto,\\ Toronto M5S 1A7, Canada} \\
\r {15} {\eightit The Johns Hopkins University, Baltimore, Maryland 21218} \\
\r {16} {\eightit Institut f\"{u}r Experimentelle Kernphysik, 
Universit\"{a}t Karlsruhe, 76128 Karlsruhe, Germany} \\
\r {17} {\eightit National Laboratory for High Energy Physics (KEK), Tsukuba, 
Ibaraki 315, Japan} \\
\r {18} {\eightit Ernest Orlando Lawrence Berkeley National Laboratory, 
Berkeley, California 94720} \\
\r {19} {\eightit Massachusetts Institute of Technology, Cambridge,
Massachusetts  02139} \\   
\r {20} {\eightit University of Michigan, Ann Arbor, Michigan 48109} \\
\r {21} {\eightit Michigan State University, East Lansing, Michigan  48824} \\
\r {22} {\eightit University of New Mexico, Albuquerque, New Mexico 87131} \\
\r {23} {\eightit The Ohio State University, Columbus, OH 43210} \\
\r {24} {\eightit Osaka City University, Osaka 588, Japan} \\
\r {25} {\eightit Universita di Padova, Istituto Nazionale di Fisica 
          Nucleare, Sezione di Padova, I-36132 Padova, Italy} \\
\r {26} {\eightit University of Pennsylvania, Philadelphia, 
        Pennsylvania 19104} \\   
\r {27} {\eightit Istituto Nazionale di Fisica Nucleare, University and Scuola
               Normale Superiore of Pisa, I-56100 Pisa, Italy} \\
\r {28} {\eightit University of Pittsburgh, Pittsburgh, Pennsylvania 15260} \\
\r {29} {\eightit Purdue University, West Lafayette, Indiana 47907} \\
\r {30} {\eightit University of Rochester, Rochester, New York 14627} \\
\r {31} {\eightit Rockefeller University, New York, New York 10021} \\
\r {32} {\eightit Rutgers University, Piscataway, New Jersey 08855} \\
\r {33} {\eightit Academia Sinica, Taipei, Taiwan 11530, Republic of China} \\
\r {34} {\eightit Texas A\&M University, College Station, Texas 77843} \\
\r {35} {\eightit Texas Tech University, Lubbock, Texas 79409} \\
\r {36} {\eightit University of Tsukuba, Tsukuba, Ibaraki 315, Japan} \\
\r {37} {\eightit Tufts University, Medford, Massachusetts 02155} \\
\r {38} {\eightit Waseda University, Tokyo 169, Japan} \\
\r {39} {\eightit University of Wisconsin, Madison, Wisconsin 53706} \\
\r {40} {\eightit Yale University, New Haven, Connecticut 06520} \\
\end{center}
}
\maketitle
\vskip -0.1in
\begin{abstract}
We present a study of events with $W$ bosons and hadronic jets produced
in \pbp\ collisions at a center of mass energy of 1.8 TeV. The data
consist of 51400 \Wenu\ decay candidates from 108~\invpb\ of integrated
luminosity collected with the CDF detector at the Tevatron Collider. The
cross sections and jet production properties have been measured for $W$
+ $\geq$1 to $\geq$4 jet events. The data are compared to predictions
of leading order QCD matrix element calculations with added gluon
radiation and simulated parton fragmentation. 
\end{abstract}
\vskip -0.1in
\pacs{
14.70.Fm, 12.38.Qk, 13.85.Qk}

\narrowtext
\twocolumn

The production of $W$ bosons in \pbp\ collisions at the Fermilab Tevatron
Collider provides the opportunity to test perturbative QCD predictions at
large momentum transfers. Previous analyses have used these data to study
$W$ production and decay properties
\cite{xsec,oldw,D0_nlo_onejet}, diboson ($WW$, $WZ$, $W\gamma$)
production \cite{diboson,W_photon}, and the pair production of top quarks
\cite{topprd,topprl}.  In this Letter, we present cross section measurements
and kinematic properties of direct single $W$ boson production with jets. 
After $W$ events from top decay are removed, the data are compared to quantum
chromodynamics (QCD)
predictions of single $W$ + jet production.  These comparisons test how well
the standard model predicts hadronic production properties of $W$ bosons
at the highest center of mass energies studied to date.

This analysis uses 108 \invpb\ of integrated luminosity collected with
the CDF detector \cite{CDF_Detectors} from 1992--95.  The principal
detector elements used for this measurement are the vertex tracking
chamber (VTX), the central tracking chamber (CTC), and the calorimeters. 
The VTX, a wire time-projection chamber, locates interactions along the
beam direction.  The CTC, a cylindrical drift 
chamber, measures the momenta of charged particles in the
region $|\eta| < 1.1$ \cite{coord}. Both tracking detectors are immersed
in a 1.4 T magnetic field. The electromagnetic
and hadronic calorimeters cover the range $|\eta| < 4.2$ and are used to
measure the energies of electrons and jets. 

\Wenu\ decay candidates are identified in events that pass a high
transverse energy (\et\ $=E\sin{\theta}$) electron trigger.  The event selection
requires an isolated \cite{isolated} electron in the central calorimeter
($|\eta| \leq 1.1$) that has \et\ $\geq$ 20~GeV and satisfies tight
selection criteria \cite{tight_cuts}.  The reconstructed neutrino transverse
energy ($\met$), measured from the imbalance of \et\ in the calorimeter,
must exceed 30~GeV.  

Jets in the $W$ events are clustered using a cone algorithm
\cite{jetclu} with radius $\Delta R \equiv \sqrt{\Delta\eta^2 + \Delta\phi^2} =
0.4$.  
We account for parton energy deposited outside the cone, and
correct for energy contaminating the cone from both the underlying event
and additional  \pbp\ interactions.
We count jets with \et\ $\geq$ 15~GeV and $|\eta| \leq
2.4$, and reject any events that have a jet within $\Delta R$ = 0.52
of an electron.  Of 51431 $W$ candidate events, 11144 events have
$\geq$1 jet, 2596 have $\geq$2 jets, 580 have $\geq$3 jets, 126 have
$\geq$4 jets, and 21 have $\geq$5 jets.  These jet multiplicities are
subsequently corrected for jets produced in additional \pbp\ interactions
that occur in the same bunch crossing as the $W$ event.  There is a 1\%
probability that an event will have a single extra jet; the probability
drops by about a factor of 6 for each additional extra jet.

The systematic uncertainties on jet counting are determined by varying 
the jet energy by
$\pm$5\%, the jet $|\eta|$ by $\pm$0.2, the probability of jets from
additional \pbp\ interactions by $^{+100\%}_{-50\%}$, and the correction
for energy contamination in the jet cone by $\pm$50\% ($\pm$0.5~GeV on average).
The combined uncertainty ranges from 10\% for the $\geq$1
jet sample to 30\% for the $\geq$4 jet sample and dominates the uncertainties in
the $W$ + jet cross section measurements. 

We measure the cross section for \pbp\ $\rightarrow W$ production as a
function of jet multiplicity $n$ from the number of observed $W$
candidates with $\geq$$n$ jets $(N_n)$ using the equation
\begin{equation}
  \sigma_n\cdot {\mathrm BR} =
  \sigma_0\cdot {\mathrm BR}
  \cdot {N_n - B_n \over \epsilon_n}
  \cdot {\epsilon_0 \over N_0 - B_0} \label{cs}
\end{equation}
where $\epsilon_{n}$ is the $W$ detection efficiency, $B_{n}$ is the
estimated background, and BR is the branching ratio for \Wenu\ decay. 
For the inclusive cross section times branching ratio ($\sigma_0
\cdot$BR) we use a previous CDF measurement of 2490 $\pm$ 120 pb
\cite{WZxsec}.  This method takes advantage of the cancellation of some
systematic uncertainties in the ratios, and gives the most
accurate relative $W$ + $\geq$$n$ jet cross sections.

The estimate of $B_{n}$ in Eq.~(\ref{cs}) includes
$Z \rightarrow e^+e^-$, $W \rightarrow \tau\nu$, and
direct QCD multijet production. The $Z \rightarrow e^+e^-$ and $W \rightarrow 
\tau\nu$
contributions are small (3\%) and have a negligible effect on the ratios.
Multijet contamination is measured with 
a sample obtained by removing the 
electron isolation
and $\met$ requirements of the $W$ selection and then 
extrapolating from the multijet-dominated region 
into the $W$ signal region. This background ranges from (2.9 $\pm$ 0.9)\% for 
the $\geq$0 jet sample to (27 $\pm$ 11)\% for the $\geq$4 jet sample.

In order to isolate direct single $W$ production
($q\overline{q'} \rightarrow W$), $B_{n}$ also includes standard model 
predictions of diboson and top quark production. The contribution from $WW$ and 
$WZ$ production is negligible; however, we include a correction to the jet
multiplicity to account for $W\gamma$ events in which the photon is
reconstructed as a jet.  This is predicted to occur in (0.4 $\pm$ 0.1)\% of
events.  The rate of \Wenu\ events from standard model $t\overline t$ ranges
from  ($0.08\pm 0.02$)\% for $\ge$0  jet events to ($26\pm 5$)\% for $\ge$4 jet
events. The cross sections and kinematic distributions are corrected for these
contributions.

The final correction to the number of \Wenu\ + $\geq$$n$ jet candidates
accounts for the efficiency ($\epsilon_n$ in Eq.~(\ref{cs}))
of identifying \Wenu\ decays.
The acceptance due to restrictions on the electron \et,\hspace{.08in}$\met$, and
detector fiducial volume was determined using a leading-order QCD
calculation \cite{VECBOS} for $W$ + $\geq$1 to $\geq$4 jets.  
The electron-jet overlap rate is calculated  directly from the
data by taking \Zee\ events and replacing the \Zee\ decay with
a simulated \Wenu\ decay, preserving the boson transverse momentum ($p_T$). 
The overall efficiency also includes the 
efficiency of the online trigger and the efficiency of the
electron identification.  The combined efficiency is
(19.6 $\pm$ 0.3)\% for $W+$ $\geq$0 jets and remains 
nearly constant as a function of the number of jets.

\begin{figure}[tb]
\begin{center}
\epsfxsize=3.4in
\epsfbox{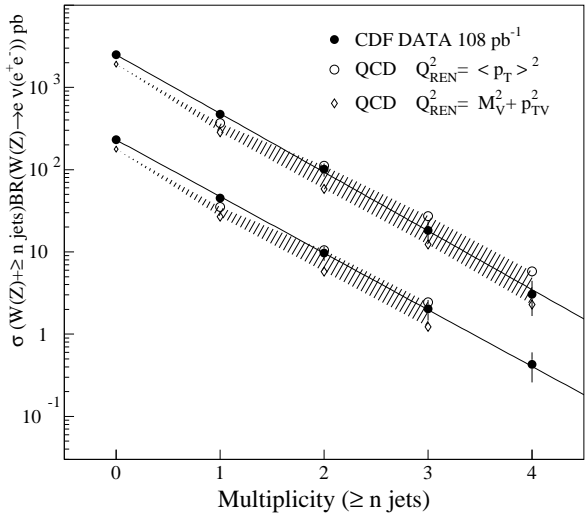}
\nobreak
{\small \caption{
Cross sections for \Wenu\ + $\geq$$n$ jets (top) and \Zee\ + $\geq$$n$ jets
(bottom) versus inclusive jet multiplicity. The lines are fits of an
exponential to the data. The theory is shown as a
shaded band which represents the uncertainty due to the renormalization scale.
The $\ge$0  jet prediction is a Born-level calculation for $W$ production.
\label{fig:xsec}}}
\end{center}
\end{figure}
The measured cross sections for single \Wenu\ + $\geq$$n$ jet events are
listed in Table~\ref{tab:results} and plotted in Fig.~\ref{fig:xsec}. 
Also included in Table~\ref{tab:results} are the ratios $\sigma_{n}
/\sigma_{n-1}$, which 
show that the cross sections fall by about a factor
of five with each additional jet.

    The measured cross sections and kinematic distributions are compared to
predictions of leading order (LO) pertubative QCD using the VECBOS
\cite{VECBOS} Monte Carlo program.  We use a two-loop $\alpha_{s}$ evolution
evaluated at renormalization scales of either \qapts\ of the partons or \qhs\
of the boson, representing reasonable extremes.  The CTEQ3M \cite{CTEQ3M}
parton density functions are used with the factorization scale set to the
renormalization scale.   The QCD predictions are at least five times more
sensitive to the renormalization scale than to the factorization scale or the
choice of parton density functions (e.g. MRSA' \cite{MRSAprime}). Initial state
gluon radiation, final state parton fragmentation, and hadronization are
simulated using the HERWIG \cite{HERWIG} Monte Carlo program.   This procedure
represents a partial higher-order correction to the tree-level diagrams and we
refer to it as enhanced leading order (ELO). The generated hadronic showers are
processed with the full CDF detector simulation. The same reconstruction and
selection criteria applied to experimental data are used on the simulated data.

\begin{figure}[tb]
\begin{center}
\epsfxsize=3.4in
\epsfbox{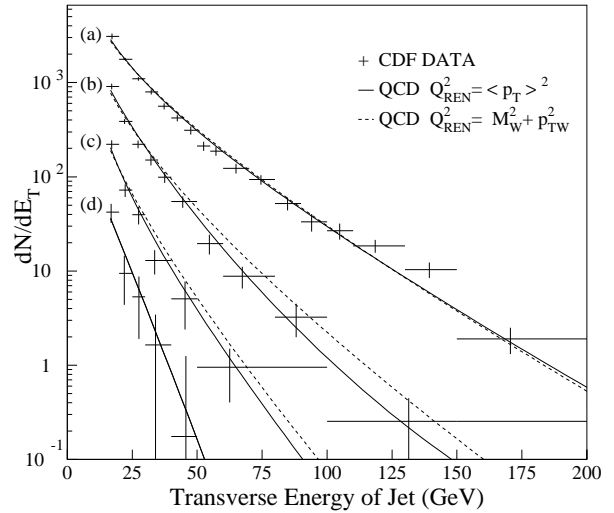}
\nobreak
{\small \caption{
Transverse energy distribution of the 
(a) highest $E_{T}$ jet in $\geq$1 jet events, (b) 
second highest $E_{T}$ jet in $\geq$2 jet events, (c) 
third highest $E_{T}$ jet in $\geq$3 jet events, and 
(d) fourth highest $E_{T}$ jet in $\geq$4 jet events.
The curves represent the ELO QCD
predictions.  The renormalization scale $Q^2$ is \apts\ for the solid curves
and \qhs\ for the dashed curves (a--c only).  The theory is normalized to the
data for each distribution and the errors are the sum of statistical and 
systematic uncertainties. 
\label{fig:Et}}}
\end{center}
\end{figure}

The QCD predictions are listed in Table~\ref{tab:results} and plotted
with the measured cross sections in Fig.~\ref{fig:xsec}.  The
sensitivity of the calculated cross sections to the renormalization
scale is indicated by the shaded band.  The harder $Q^2$ scale (\qhs)
predicts relative cross sections that are consistent with the measured
cross sections but are low in magnitude by about a factor 1.6, while the softer
$Q^2$ scale (\apts) predicts cross sections generally closer in
magnitude but with a ratio ranging from 1.28 for $\ge$1 jets 
to 0.53 for $\ge$4 jets.
Thus, within the inherent uncertainty of the LO
calculation, the predicted and measured $W$ + $\geq$$n$ jet cross
sections are in agreement for $n$ = 2 to 4.  For comparison,
Fig.~\ref{fig:xsec} also shows the cross sections and QCD predictions
for $Z$ + $\geq$$n$ jets from a previous CDF measurement \cite{prl_z}. 
These have the same general features as $W$ production, but are lower in
cross section by about a factor of 10.

\begin{figure}[tb]
\begin{center}
\epsfxsize=3.4in
\epsfbox{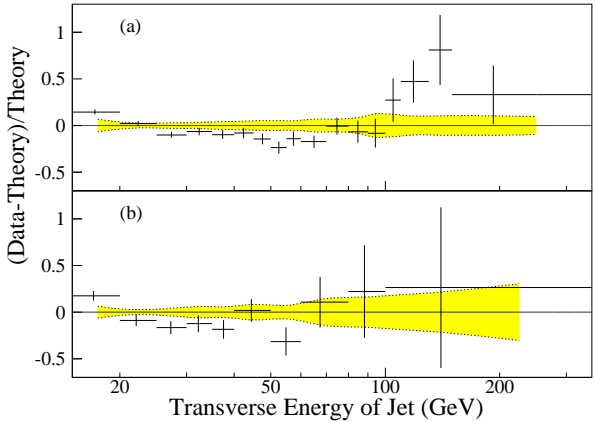}
\nobreak
{\small \caption{
(Data--theory)/theory (\qapts) for the jet transverse energy distribution 
of the first
and second highest $E_{T}$ jets in (a) $\geq$1 and (b) $\geq$2 jet
events, respectively.  The error bars are statistical uncertainties and
the band represents the systematic uncertainty on the shape.
The theory is normalized to the data.
\label{fig:DMT_Et}}}
\end{center}
\end{figure}

\begin{figure}[tb]
\begin{center}
\epsfxsize=3.4in
\epsfbox{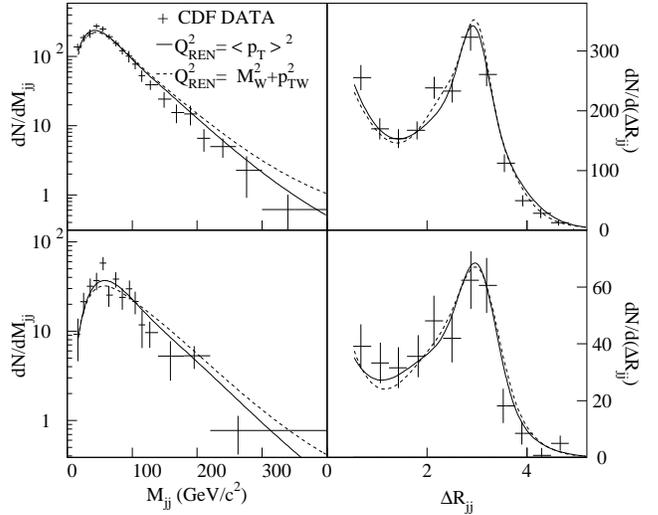}
\nobreak
{\small \caption{
Distributions of dijet mass and separation in $\eta-\phi$ space between the two
highest-\et\ jets for $W$ + $\geq$2 jet events (top) and $W$ + $\geq$3
jet events (bottom).  The curves represent the ELO QCD
predictions: $Q^2=$ \apts\ (solid) and $Q^2=$ \qhs\ (dashed). 
\label{fig:jj_corr}}}
\end{center}
\end{figure}

Details of the QCD predictions are studied using kinematic distributions
of jets in $W$ events. We select events from the ELO simulated $W$ sample
with the same selection criteria used for the data. Shape comparisons 
are made by normalizing the theory to data.
We first compare
the measured \et\ spectra (Fig.~\ref{fig:Et}) of jets 1--4, ordered by
decreasing \et, to the ELO QCD prediction \cite{fit_descr}.  The
sensitivity of the prediction to the renormalization scale is
illustrated by varying it from \apts\ (solid curve) to \qhs\ (dotted
curve).  The correspondence 
between data and theory for these
distributions is more clearly seen in  Fig.~\ref{fig:DMT_Et}, which shows
(data--theory)/theory for the same spectra.  Correlations between jets
are studied by measuring the separation (\drjj) and invariant mass
(\mjj) of pairs of jets.  The distributions of \drjj\ and \mjj\ for the
two highest \et\ jets in $\geq$2 jet and $\geq$3 jet events are shown in
Fig.~\ref{fig:jj_corr}.  The systematic uncertainties
are determined by the change in the distributions when the jet energy 
and subtracted backgrounds are varied independently within their $\pm
1\sigma$ limits.  The error
bars in Figs.~\ref{fig:Et} and \ref{fig:jj_corr} include statistical
and systematic uncertainties.

The shape comparisons in Figs.~\ref{fig:Et}--\ref{fig:jj_corr}
demonstrate that the ELO QCD predictions reproduce the main
features of both the jet \et\ and jet-jet correlation distributions.  In
particular, 
the measured and predicted jet \et\ spectra for the four 
highest \et\ jets generally remain within 15\% over three orders of
magnitude.  The correlation between jets, as measured by \drjj, is well
predicted by the QCD calculation (Fig.~\ref{fig:jj_corr}c and d), and
the measured invariant mass distributions (Fig.~\ref{fig:jj_corr}a and
b) are in fair agreement with the QCD predictions.  However, the high
statistics of our $W$ + $\geq$1 jet sample show the limitation of this
QCD prediction.   Fig.~\ref{fig:DMT_Et}a shows that the theory calculation
underestimates the cross section for the lowest \et\ ($<20$~GeV) and highest
\et\ ($>100$~GeV) jets.  These regions rely heavily upon the partial
higher-order corrections generated by HERWIG.  At low \et, initial state gluon
radiation is sometimes hard enough to become the 
highest \et\ jet and supersede
the parton generated in the LO matrix element.  For events with the highest jet
\et\ $>100$~GeV, over 50\% of the $W$ + $\geq$1 jet events have at least 2 jets
which explicitly indicates the need for higher-order corrections to the $W$ + 1
parton calculation.  As expected, the ELO QCD calculation only partly corrects
for the higher-order QCD terms.

In summary, this Letter contains an analysis of jet production
associated with \Wenu\ events selected from 108 \invpb\ of
\pbp\ collisions at a center of mass energy of 1.8 TeV.  Data are
compared to enhanced LO QCD predictions (LO parton matrix elements with
HERWIG-simulated fragmentation) to determine the reliability of QCD
calculations.  The ratio of the measured to predicted cross section is
1.28 $\pm$ 0.16 (\qapts) and 1.65 $\pm$ 0.20 (\qqhs) for 
\pbp\ $\rightarrow W$ + $\geq$1 jet events.
For higher jet
multiplicities, the two $Q^2$ predictions bracket the measurement, with
the \qqhs\ prediction at a approximately constant fraction
below the measured cross section.  The shapes of the QCD-predicted jet
production properties are in general agreement with the data, but the
statistics of the \Wenu\ data are large enough to show some limitations
of the enhanced LO QCD predictions.

\begin{table}[t]
\nopagebreak
\widetext
\begin{center}
\begin{tabular}{ccccccc}
 $n$ &
 $\sigma_{n} \cdot {\rm BR}(W^\pm \rightarrow e^\pm\nu)$ &
 \multicolumn{2}{c}{\qapts} &
 \multicolumn{2}{c}{\qqhs} &
 $\sigma_{n}/\sigma_{n-1}$ \\
 Jets &
 (pb) &
 $BR \cdot \sigma_{\rm QCD}$ &
 $\sigma_{\rm Data}/\sigma_{\rm QCD}$ &
 $BR \cdot \sigma_{\rm QCD}$ &
 $\sigma_{\rm Data}/\sigma_{\rm QCD}$ &
 (Data) \\
\hline

$\geq$1  &   471  $\pm$ 5  $\pm$ 57  
&  367 $\pm$ 5  &  1.28 $\pm$ 0.16
&  285 $\pm$ 4 &  1.65 $\pm$ 0.20
&  0.189 $\pm$ 0.021 \\

$\geq$2  &   101  $\pm$ 2  $\pm$ 19  
& 112 $\pm$ 5    &  0.90 $\pm$ 0.17
&  58.1 $\pm$ 1.5  &  1.74 $\pm$ 0.33
&  0.214 $\pm$ 0.015 \\

$\geq$3  &   18.4  $\pm$ 1.1  $\pm$ 5.2  
&  27.2 $\pm$ 2.1  &  0.67 $\pm$ 0.20
&  12.3 $\pm$ 0.6  &  1.49 $\pm$ 0.44
&  0.182 $\pm$ 0.020 \\

$\geq$4  &   3.1  $\pm$ 0.6  $\pm$ 1.3  
&  5.8 $\pm$ 0.7  & 0.53 $\pm$ 0.25
&  2.3 $\pm$ 0.2  & 1.33 $\pm$ 0.62
&  0.166 $\pm$ 0.042 \\

$\geq$5  &  0.24  $\pm$ 0.24  $\pm$ 0.28  
& & & &
&  0.080 $\pm$ 0.109 \\
\end{tabular}
\caption{
$W$ + $\geq$$n$ jet cross sections.  The first error on the data cross
sections is the statistical error; the second includes the systematic
error on the $W$ acceptance, the background systematic, and the
jet-counting uncertainty, as described in the text.  The QCD Monte Carlo
cross sections ($BR \cdot \sigma_{\rm QCD}$) are generated using 
VECBOS for $Q^2$ scales of \apts\ and
\qhs.  The uncertainties for the theory are statistical.
\label{tab:results}}
\end{center}\narrowtext
\end{table}

We thank the Fermilab staff and the technical staffs of the
participating institutions for their vital contributions.  We also thank
Walter Giele and Nigel Glover for many useful discussions. This
work was supported by the U.S. Department of Energy and National Science
Foundation; the Italian Istituto Nazionale di Fisica Nucleare; the Ministry of
Education, Science and Culture of Japan; the Natural Sciences and Engineering
Research Council of Canada; the National Science Council of the Republic of
China; and the A. P. Sloan Foundation.


\begin{thebibliography}{10}
\bibitem{coord}
We use a coordinate system in which $z$ is along the proton
direction, $\phi$ is the azimuthal angle, $\theta$ is the polar
angle and $\eta \equiv - \ln (\tan {\frac{\theta}{2}})$ is
the pseudorapidity. We take $z$=0 at the center of the detector for fiducial
cuts and at the interaction point for event variables.
\bibitem{isolated}
An isolated electron is one for which the calorimeter \et\ in a cone of
radius 0.4 in $\eta$--$\phi$ around the electron cluster is less than
10\% of the electron \et.
\bibitem{fit_descr}
The curves are fits of analytic functions to the QCD Monte Carlo prediction
which accurately reproduce the spectra.
\end{thebibliography}
\end{document}